\pdfoutput=1

\documentclass[11pt,twoside,a4paper,cmspaper,final,collab]{cms-tdr}
\def\svnVersion{420985P}\def\cmsCernNoTag{CERN-EP-2016-098}\def\cmsCernDate{\today}\def\cmsMessage{Published in Physics Letters B as \href{http://dx.doi.org/10.1016/j.physletb.2017.07.001}{\doi{10.1016/j.physletb.2017.07.001.}}}

\begin{document}\cmsNoteHeader{HIN-12-009}

\hyphenation{had-ron-i-za-tion}
\hyphenation{cal-or-i-me-ter}
\hyphenation{de-vices}
\RCS$Revision: 279807 $
\RCS$Id: HIN-12-009.tex 279807 2015-03-06 23:38:25Z tapiata $
\newlength\cmsFigWidth
\ifthenelse{\boolean{cms@external}}{\setlength\cmsFigWidth{\columnwidth}}{\setlength\cmsFigWidth{0.6\textwidth}}
\ifthenelse{\boolean{cms@external}}{\providecommand{\cmsLeft}{top}}{\providecommand{\cmsLeft}{left}}
\ifthenelse{\boolean{cms@external}}{\providecommand{\cmsRight}{bottom}}{\providecommand{\cmsRight}{right}}
\providecommand{\Pb}{\ensuremath{\mathrm{Pb}}\space}
\providecommand{\STARLIGHT}{\textsc{starlight}\xspace}
\providecommand{\Xznz}{\ensuremath{X_{\mathrm{n}}0_{\mathrm{n}}}\xspace}
\providecommand{\XnXn}{\ensuremath{X_{\mathrm{n}}X_{\mathrm{n}}}\xspace}
\providecommand{\Xnzn}{\ensuremath{X_{\mathrm{n}}0_{\mathrm{n}}}\xspace}

\cmsNoteHeader{HIN-12-009}
\title{Coherent \JPsi photoproduction in ultra-peripheral PbPb collisions at $\sqrt{s_{NN}} = 2.76$\TeV with the CMS experiment}

\date{\today}

\abstract{
The cross section for coherent \JPsi photoproduction accompanied by at least one neutron on one side of the interaction point and no neutron activity on the other side, $\Xznz$, is measured with the CMS experiment in ultra-peripheral PbPb collisions at $\sqrt{s_{NN}} = 2.76$\TeV. The analysis is based on a data sample corresponding to an integrated luminosity of 159\mubinv, collected during the 2011 PbPb run. The \JPsi mesons are reconstructed in the dimuon decay channel, while neutrons are detected using zero degree calorimeters. The measured cross section is $\rd\sigma^{\text{coh}}_{\Xznz}/{\rd y}(\JPsi)  = 0.36\pm0.04\stat\pm0.04\syst\unit{mb}$ in the rapidity interval $1.8<\abs{y}<2.3$. Using a model for the relative rate of coherent photoproduction processes, this $\Xznz$ measurement gives a total coherent photoproduction cross section of ${\rd\sigma^{\text{coh}}}/{\rd y}(\JPsi)  = 1.82\pm 0.22\stat\pm0.20\syst\pm0.19\thy\unit{mb}$. The data strongly disfavour the impulse approximation model prediction, indicating that nuclear effects are needed to describe coherent \JPsi photoproduction in $\gamma + \Pb$ interactions. The data are found to be consistent with the leading twist approximation, which includes nuclear gluon shadowing.}

\hypersetup{%
pdfauthor={CMS Collaboration},%
pdftitle={Coherent J/Psi photoproduction in ultra-peripheral PbPb collisions at sqrt(s[NN]) = 2.76 TeV with the CMS experiment},%
pdfsubject={CMS},%
pdfkeywords={CMS, physics, heavy ion collisions, ultra-peripheral collisions, UPC, J/Psi}}

\maketitle

\section{\label{sec:Intro} Introduction}
Photon-induced reactions are dominant in Ultra-Peripheral Collisions (UPC) of heavy ions, which involve electromagnetic interactions at large impact parameters of the colliding nuclei. Because of the extremely high photon flux in ultra-peripheral heavy-ion collisions which is proportional to $Z^2$, where $Z$ is the charge of the nucleus, photon-nucleus collisions at the LHC are abundant~\cite{Bertulani:2005ru, Baltz:2007kq,Contreras:2015dqa}. Furthermore, in UPCs the LHC can reach unprecedented photon-lead and photon-proton center-of-mass energies.

Vector meson photoproduction in UPCs has received recent interest~\cite{Contreras:2015dqa}. Exclusive \JPsi photoproduction off protons is defined by the reaction $\gamma + \Pp  \to \JPsi + \Pp$, with the characteristic features that, apart from the vector meson in the final state, no other particles are produced and the vector meson has a mean transverse momentum significantly lower than in inclusive reactions. Another characteristic feature is that in exclusive photoproduction the quantum numbers of the final state can be studied unambiguously. The $\gamma + \Pp \to \JPsi + \Pp$ production process has been studied by H1 and ZEUS collaborations at the electron-proton collider HERA~\cite{Alexa:2013xxa,Aktas:2005xu,Chekanov:2002xi}, by the CDF collaboration in proton-antiproton collisions at the Tevatron~\cite{Aaltonen:2009kg}, and by the ALICE and LHCb collaborations at the LHC, in proton-lead~\cite{TheALICE:2014dwa} and proton-proton collisions~\cite{Aaij:2014iea}, respectively. Since the cross section of photoproduced vector mesons such as \JPsi, $\Pgy$, and $\Upsilon \mathrm{(nS)}$, in leading order perturbative QCD, is proportional to the gluon density squared in the target~\cite{Ryskin:1992ui, Brodsky:1994kf}, the study of such diffractive processes in high-energy collisions is expected to provide insights into the role played by gluons in hadronic matter. As an example, a \JPsi produced at rapidity $y$ is sensitive to the gluon distribution at 
$x = ({M_{\JPsi}}/{\sqrt{s}})\re^{\pm y}$ at hard scales $Q^2\sim {M_{\JPsi}^2}/{4}$, where $M_{\JPsi}$ is the \JPsi mass, $\sqrt{s}$ is the 
center-of-mass energy of the colliding system and $y$ is the rapidity of the \JPsi~\cite{Ryskin:1992ui, Brodsky:1994kf}. The relevant values 
of $x$ that can be explored in this analysis are in the $10^{-2}$ to $10^{-5}$ range.

In ultra-peripheral nucleus-nucleus collisions, vector mesons can be produced in $\gamma + \mathrm{A}$ interactions off one of the nuclei~\cite{Adeluyi:2012ph,Adeluyi:2013tuu,Cisek:2012yt,Klein:1999qj,Lappi:2013am,Goncalves:2011vf,Rebyakova:2011vf,Guzey:2013jaa,Ivanov:2007ms}. Such interactions are characterized by very low multiplicity, and indeed the majority of such events are exclusive, \ie $\gamma + \mathrm{A}  \to \JPsi+\mathrm{A} $. The interaction that produces the vector meson is classified as coherent if the photon interacts with the whole nucleus, leaving the nucleus intact. In incoherent interactions, the photon interacts with a single nucleon, and the nucleus breaks apart. The requirement of having coherent photoproduction constrains the mean transverse momentum of the vector mesons to be of the order of $\pt \approx 60\MeV$ for PbPb collisions at $\sqrt{s_{NN}} = 2.76$\TeV~\cite{Bertulani:2005ru, Baltz:2007kq,Contreras:2015dqa}. This follows from the fact that the transverse momentum distribution is driven by the target form factor. Because the nucleon radius is smaller than that of the nuclei, the momentum transfer to the vector meson from incoherent photoproduction is higher, of the order of 500\MeV at $\sqrt{s_{NN}} = 2.76$\TeV. Such a momentum transfer causes the target nucleus to break up and, in most cases, it produces neutrons at very small angles with respect to the Pb beams (forward neutrons). However, vector mesons produced coherently can also be accompanied by forward neutrons. Owing to the intense electromagnetic fields present in ultra-peripheral nucleus-nucleus collisions, additional independent soft electromagnetic interactions can occur between the nuclei giving rise to forward neutrons. The emission of such neutrons is understood in terms of giant dipole resonances~\cite{Berman:1975tt}. Neutron-differential studies are considered as a promising tool to decouple low-$x$ and high-$x$ contributions in vector meson photoproduction, \eg~\cite{Guzey:2016piu}. 

Ultimately, UPC studies at hadron colliders and similar measurements at the proposed electron-ion colliders~\cite{Accardi:2012qut,AbelleiraFernandez:2012cc} are expected to reduce uncertainties in our knowledge of the initial state of a high-energy nucleus-nucleus collision, in particular, regarding the intrinsic distribution and fluctuations of gluons in the nuclei. The uncertainty over the initial state is currently an impediment to measuring fundamental properties of the quark-gluon plasma, such as viscosity, to a high precision~\cite{Akiba:2015jwa}. The largest theoretical uncertainty comes from the gluon distribution function in nuclei, which at a given value of the Bjorken variable $x$ may be depleted (shadowing) or enhanced (anti-shadowing) with respect to the scaled gluon distribution function in the proton. These parton distribution functions (PDFs) have been parameterized using global fitting techniques, such as EPS09~\cite{Eskola:2009uj}, that evolve quark, antiquark, and gluon distributions as a function of $Q^{2}$. The fitting results from EPS09 have a large uncertainty for gluon PDFs for $x < 10^{-2}$ and low $Q^{2}$ due to the lack of experimental data. The data from ultra-peripheral collisions at the LHC have the potential to provide new constraints to the gluon PDFs in protons and nuclei. Recent theoretical work has been carried out to include the study of UPC vector meson photoproduction in global PDF fits~\cite{Jones:2015nna,Rojo:2015acz}.

The STAR and PHENIX collaborations at RHIC have studied $\rho^{0}$ and \JPsi photoproduction in ultra-peripheral AuAu collisions~\cite{Abelev:2007nb, Afanasiev:2009hy,Agakishiev:2011me}. Although RHIC studies have demonstrated the feasibility of measuring these processes, it was not possible to significantly constrain the nuclear gluon PDFs. The \JPsi analysis was statistically limited~\cite{Abelev:2007nb}, while for UPC $\rho^{0}$ analyses a hard scale cannot be established to perform perturbative QCD calculations. The production rate for UPC physics processes is much higher at the LHC. The ALICE collaboration has measured coherent photoproduction of \JPsi mesons in ultra-peripheral PbPb collisions at $\sqrt{s_{NN}} = 2.76$\TeV~\cite{Abelev:2012ba, Abbas:2013oua}. These data have been used to compute the nuclear suppression
factor $R = (G_{A}/AG_{N})^{2}$, where $G_{A}$ and $G_{N}$ are the gluon distributions in a nucleus ($A = 208$ in the case of the Pb nuclei) and in a free proton, respectively, obtaining $R = 0.61^{+0.05}_{-0.04}$ for $x \sim 10^{-3}$~\cite{Guzey:2013xba}. These results have provided evidence that the nuclear gluon density is below that expected for a simple superposition of protons and neutrons in the nucleus~\cite{Abelev:2012ba, Abbas:2013oua}. Models that neglect nuclear gluon shadowing such as \STARLIGHT~\cite{Klein:2016yzr} and the impulse approximation~\cite{Guzey:2013jaa}, or models that maximize the gluon shadowing, such as EPS08~\cite{Eskola:2008ca}, have been ruled out by these measurements.

This Letter reports the study of the coherent \JPsi photoproduction cross section measured in ultra-peripheral PbPb collisions at $\sqrt{s_{NN}} = 2.76$\TeV, as well as the dependence of this cross section on the associated production of forward or backward neutrons, \ie on the so-called neutron break-up mode ratios~\cite{Rebyakova:2011vf}. To focus on events with low backgrounds, following the experience at RHIC~\cite{Afanasiev:2009hy}, the UPC trigger selected events with at least one neutron in either the forward or backward direction from the interaction point using zero degree calorimeters. Using this trigger, both coherent and incoherent \JPsi mesons and $\gamma + \gamma \to \PGmp\PGmm$ events in conjunction with at least one neutron can be studied. This data sample is then used to measure the cross section for coherent \JPsi photoproduction accompanied by at least one neutron from soft independent processes. The \JPsi candidates are reconstructed through the dimuon decay channel in the rapidity interval $1.8<\abs{y}<2.3$, adding a new rapidity range to recent measurements of coherent \JPsi photoproduction at the LHC~\cite{Abelev:2012ba, Abbas:2013oua}.

This paper is organized as follows: Section 2 describes the CMS detector, Section 3 reports on the event selection and analysis strategy, Section 4 describes the signal extraction and corrections, Section 5 summarizes the uncertainties of the measurement, and Section 6 discusses the results. Finally, in Section 7 the summary is given.

\section{\label{sec:detector} The CMS detector}
The central feature of the CMS apparatus is a superconducting solenoid of 6\unit{m} internal diameter, providing a magnetic field of 3.8\unit{T}. Within the solenoid volume are a silicon pixel and strip tracker, a lead tungstate crystal electromagnetic calorimeter (ECAL), and a brass and scintillator hadron calorimeter (HCAL), each composed of a barrel and two endcap sections. The silicon tracker measures charged particles within the pseudorapidity range $\abs{\eta}< 2.5$. It consists of 1440 silicon pixel and 15\,148 silicon strip detector modules and is located in the 3.8\unit{T} field of the superconducting solenoid. The pseudorapidity coverage for the ECAL and HCAL detectors is $\abs{\eta}< 3.0$. Muons are measured using the CMS detector in the pseudorapidity range $\abs{\eta}< 2.4$. The muon detection planes are made using three technologies: drift tubes, cathode strip chambers, and resistive plate chambers. 
The \pt of the muons matched to reconstructed tracks is measured with a resolution better than 1.5\%~\cite{Chatrchyan:2012xi}. The Hadronic Forward (HF) calorimeters ($3.0<|\eta| < 5.2 $) complement the coverage provided by the barrel and endcap detectors. The beam scintillator counters (BSCs) are plastic scintillators that partially cover the face of the HF calorimeters. They have a pseudorapidity range between 3.9 and 4.4, with a time resolution of 3 ns. The zero degree calorimeters (ZDCs) are two \v{C}erenkov calorimeters composed of alternating layers of tungsten and quartz fibers, situated in between the two proton beam lines. They are sensitive to neutrons and photons with $|\eta|>8.3$. The HF, BSC and ZDC systems each consist of two detectors at either side of the interaction point: HF$^{\pm}$, BSC$^{\pm}$, ZDC$^{\pm}$, respectively. A more detailed description of the CMS detector, together with a definition of the coordinate system used and the relevant kinematic variables, can be found in~\cite{Chatrchyan:2008zzk}.

\section{\label{sec:DataAndAnalysis1} Event selection and Monte Carlo samples}
This analysis uses the data sample collected with the CMS detector in the 2011 PbPb run, which corresponds to an integrated luminosity of 159\mubinv~\cite{CMS:2012rua}. The events are selected with a dedicated trigger designed to record UPC \JPsi vector mesons and $\gamma + \gamma \to \PGmp\PGmm$ events. The UPC trigger has the following requirements:  an energy deposit consistent with at least one neutron in either of the ZDCs; no activity in at least one of the BSC$+$ or BSC$-$ scintillators; the presence of at least one single muon without a \pt threshold requirement, and at least one track in the pixel detector. The first three trigger requirements are implemented in hardware, while the last requirement is carried out by the software trigger. To reject beam-gas interactions and suppress non-UPC events the following requirements are imposed offline. The $z$ position of the primary vertex is required to be within 25 cm of the beam spot centre. The length of the pixel clusters must be consistent with tracks originating from this vertex. This requirement removes beam--background events that produce elongated pixel clusters. In addition, events are rejected if the time difference between two hits from the BSCs is above 20 ns with respect to the mean flight time between them (73 ns). This requirement removes beam-halo events, while keeping all the ultra-peripheral PbPb events.

As mentioned above, one of the UPC trigger requirements is the presence of at least one neutron. The events studied in this analysis are classified by the pattern of neutron deposition measured in the ZDCs~\cite{Grachov:2006ke,Chatrchyan:2012mb,CMS:2013bza}. The ZDC energy spectrum shows a clear one neutron peak and the detectors have an energy 
resolution of about 20\% for single neutrons in PbPb collisions at $\sqrt{s_{_\mathrm{NN}}} = 2.76$\TeV~\cite{Grachov:2006ke,Chatrchyan:2012mb,CMS:2013bza}. This resolution allows a good separation between events with zero, one, or multiple neutrons in a given ZDC detector. A given event is considered to have no neutrons in the ZDC if the calorimeter energy is less than 420 GeV, one neutron if the energy lies between 420 GeV and 1600 GeV, and more than one neutron if the energy is above 1600 GeV. The coherent \JPsi cross section is measured for the case when the \JPsi mesons are accompanied by at least one neutron on one side of the interaction point and no neutron activity on the other side ($\Xznz$). The $\Xznz$ break-up mode, which is conventionally written as $\Pb + \Pb\to \Pb + \Pb + $ \JPsi ($\Xznz$), is a subset of the triggered events. This break-up mode is well suited for rejecting non-UPC background due to its asymmetric configuration~\cite{Chiu:2001ij}.

Apart from the $\Xznz$ break-up mode, the UPC trigger also selects the $\XnXn$, $1_{\mathrm{n}}0_{\mathrm{n}}$, and $1_{\mathrm{n}}1_{\mathrm{n}}$ break-up modes. The $\XnXn$ mode requires that both ZDCs record at least one neutron. The $1_{\mathrm{n}}0_{\mathrm{n}}$ mode requires that one of the ZDCs detects exactly one neutron with no neutron activity on the other ZDC side. Finally, the $1_{\mathrm{n}}1_{\mathrm{n}}$ mode requires both ZDCs to have exactly one neutron.

In addition to the ZDC requirement, two selections are applied to reject non-UPC events. First, only events with exactly two reconstructed tracks are kept. 
Second, the HF cell with the largest energy deposit is required to have an energy below 3.85\GeV. This requirement, which is determined studying events 
triggered on empty bunches, ensures that the HF energy is consistent with the presence of photon-induced processes which leave very low signal in both the 
HF+ and HF$-$ detectors.

In this analysis, both muons have to satisfy the quality criteria described below, and must lie within the phase space region $1.2 < \abs{\eta} < 2.4$ and $1.2 < \pt < 1.8$\GeV. This phase space region is chosen to ensure good statistical precision on the data-driven measurement of the single-muon efficiency (see Section~\ref{sec:DataAndAnalysis2}). The CMS collaboration has developed several types of muon identification~\cite{Chatrchyan:2012xi}. In this analysis, all tracks in the silicon tracker that are identified as muons, based on information of the muon detectors, are used. The algorithm extrapolates each reconstructed silicon track outward to its most probable location within each detector of interest (ECAL, HCAL, muon system). This procedure enables the identification of single muons with very low transverse momenta. To reduce additional muons or charged particle tracks that can be misidentified as muons and to ensure good-quality reconstructed tracks, the single muons are required to pass the following criteria: more than 4 hits in the tracker, at least one of which is required to be in a pixel layer, a track fit with a  $\chi^2$ per degree of freedom less than three, and a transverse (longitudinal) impact parameter of less than 0.3 (20)\unit{cm} from the measured vertex. For this analysis, only events with dimuons having $\pt<1.0$\GeV, in the rapidity interval $1.8<\abs{y}<2.3$, are considered. The dimuon candidates are required to be within the invariant mass region $2.6<m(\PGmp\PGmm)<3.5$\GeV. No like-sign dimuon pairs are found in this region. Applying the muon quality requirements, after all other analysis selections, only rejects one dimuon candidate out of 518 events.

In order to compute acceptance and efficiency corrections and for signal extraction purposes, Monte Carlo (MC) samples for coherent \JPsi, incoherent \JPsi and $\gamma + \gamma$ events in the dimuon decay channel are generated, using the \STARLIGHT MC event generator~\cite{Klein:2016yzr,starlight,Klein:1999qj, Baltz:2009jk}. These events are processed with the full CMS simulation and reconstruction software. The \STARLIGHT generator models two-photon and photon-hadron interactions at ultra-relativistic energies. In the case of photon-nuclear reactions, it models both coherent and incoherent events using the vector meson dominance model. It uses the Glauber approach for calculating hadron-nucleus cross sections from hadron-nucleon ones, and makes use of exclusive \JPsi photoproduction in $\gamma + \Pp$ results from HERA to compute the coherent \JPsi cross section in $\gamma + \Pb$ interactions~\cite{Klein:1999qj}. The \STARLIGHT generator is also used to simulate the various break-up modes for one or both Pb nuclei, which assumes that the probabilities for exchange of multiple photons in a single event factorize in impact parameter space~\cite{Adam:2015gsa}.

\section{\label{sec:DataAndAnalysis2} Signal extraction and corrections}
\begin{figure}
\centering
\includegraphics[width=0.48\textwidth]{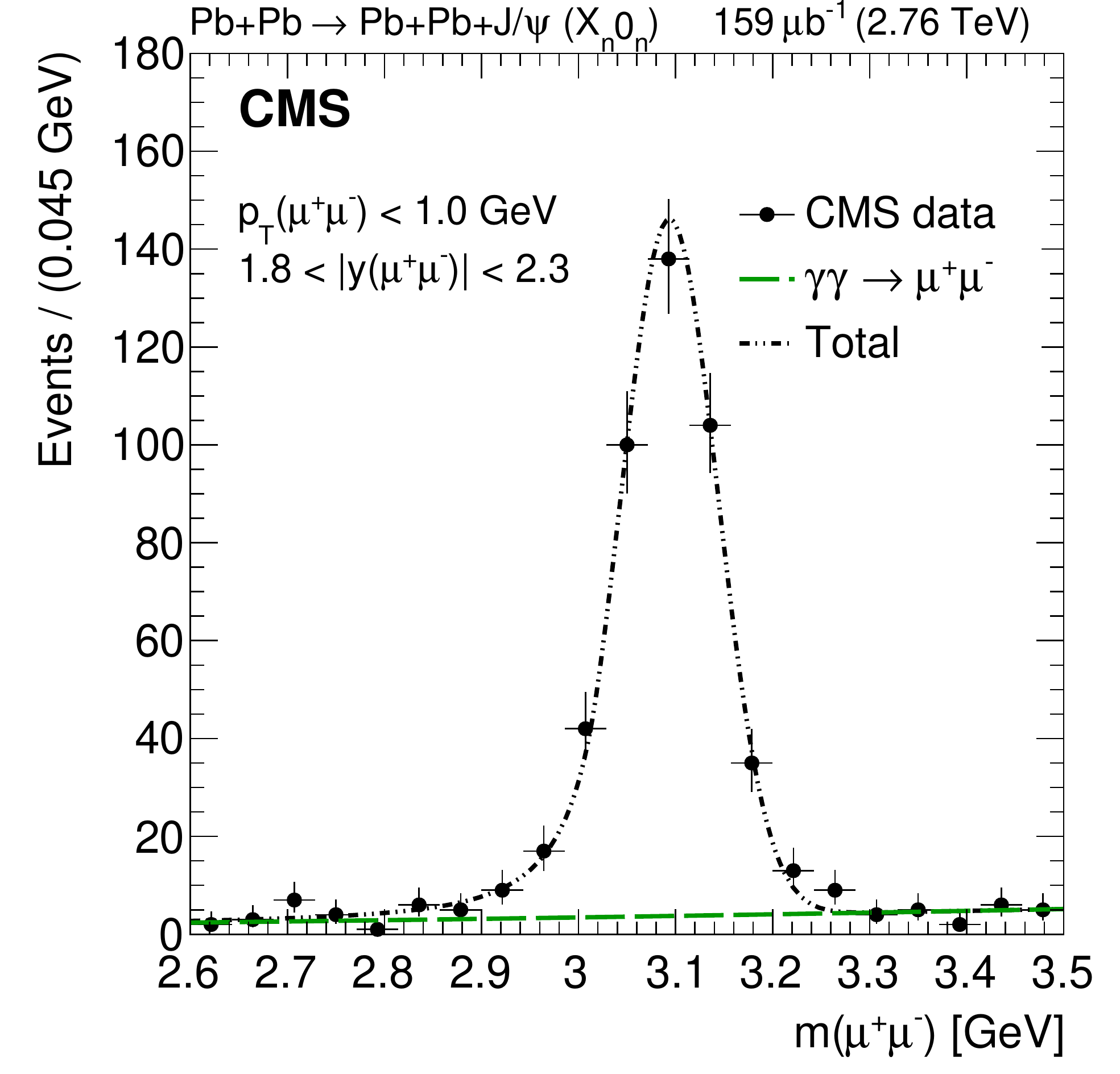}
\includegraphics[width=0.48\textwidth]{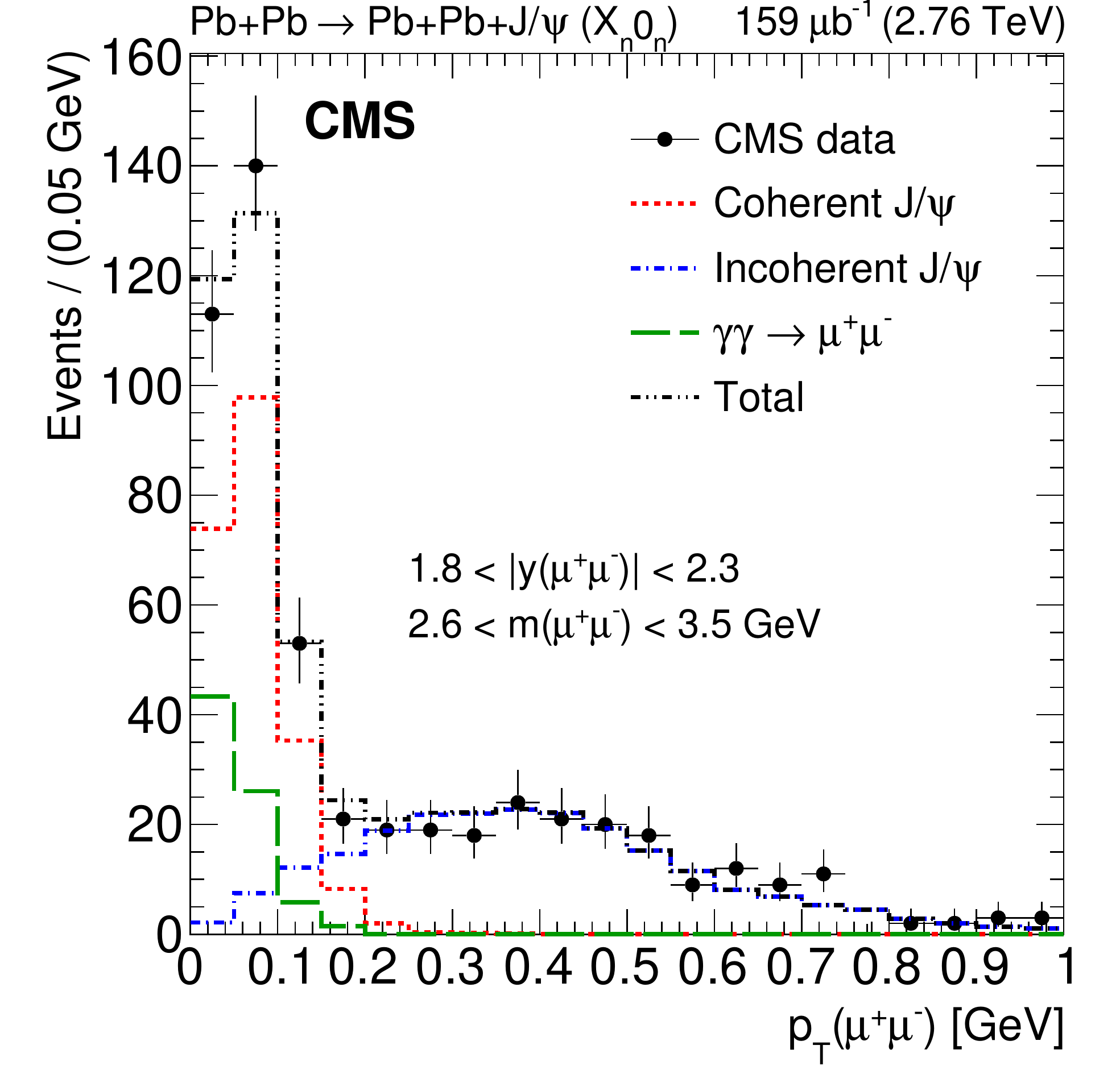}
   \caption{\label{Figure1} Results from the simultaneous fit to dimuon invariant mass (\cmsLeft) and \pt (\cmsRight) distributions from opposite-sign muon pairs with $\pt<1.0\GeV$, $1.8<\abs{y}<2.3$ and $2.6<m(\PGmp\PGmm)<3.5$\GeV for the $\Xznz$ break-up mode, after all selections are applied. In the left panel the green curve represents the $\gamma + \gamma$ component (second-order polynomial) and the black curve the sum of the $\gamma + \gamma$, incoherent \JPsi, and coherent \JPsi components (see text for details). In the right panel the green, red, and blue curves represent $\gamma + \gamma$, coherent \JPsi, and incoherent \JPsi components, respectively. The black curve represents the sum of the $\gamma + \gamma$, coherent \JPsi, and incoherent \JPsi components. Only statistical uncertainties are shown. The data are not corrected by acceptance and efficiencies, and the MC templates are folded with the detector response simulation.}
\end{figure}

After applying the selections described in Section~\ref{sec:DataAndAnalysis1}, the dimuon invariant mass and \pt distributions are simultaneously fitted in order to extract the number of coherent \JPsi, incoherent \JPsi, and $\gamma + \gamma \to \PGmp\PGmm$ events. The fit uses a maximum likelihood algorithm that takes unbinned projections of the data in invariant mass and \pt as inputs. The shapes of the \pt distributions for these three processes are determined from \STARLIGHT simulation. The yield for each of these processes in the $\pt$ distribution is a free parameter of the fit. The dimuon invariant mass distribution of the sum of coherent and incoherent \JPsi events is described with a Crystal Ball function~\cite{CrystallBall}, which accounts for the detector resolution as well as the radiative tail from internal bremsstrahlung. A second-order polynomial accounts for the underlying dimuon continuum that originates from $\gamma + \gamma \to \PGmp\PGmm$ events. The fit has nine free parameters:  three for the yields of each of the processes, two for the shape of the Crystal Ball function tail, two for the mean and width of the Crystal Ball function, and two parameters for the shape of the second-order polynomial. The fit constrains the number of coherent \JPsi, incoherent \JPsi, and dimuon continuum events to be the same in the invariant mass and \pt distributions. The projections of the $\Xznz$ break-up data onto the dimuon invariant mass and \pt axes are shown in Fig.~\ref{Figure1}. As discussed in Section~1, the average \pt distribution for the coherent events is peaked at lower \pt values than those from incoherent events. Reconstructed coherent \JPsi events are dominant for $\pt<0.15\GeV$, whereas reconstructed incoherent \JPsi events are dominant for $\pt>0.15\GeV$. For events with $\pt<0.15$\GeV and in the rapidity interval $1.8<\abs{y}<2.3$, the fit yields $207\pm18\stat$ for the coherent \JPsi candidates, $75 \pm13\stat$ for incoherent \JPsi events and $75 \pm13\stat$ for $\gamma + \gamma$ events.

In addition, the data sample is studied in terms of the following two cases: (i) neutrons emitted in the same rapidity hemisphere as the \JPsi, and (ii) neutrons emitted in the opposite rapidity hemisphere than the \JPsi. The number of coherent \JPsi events is found to be consistent, within the statistical and systematic uncertainty, between the two 
cases. This suggests that the emitted neutrons and the photoproduced \JPsi events are independent processes, within the current uncertainty. On the other hand, for incoherent \JPsi 
photoproduction most of the events are found in the configuration where the neutrons and the \JPsi mesons are produced in the same hemisphere. This suggests that in incoherent \JPsi photoproduction the low-$x$ and high-$x$ contributions are decoupled and can be more easily observed than in coherent \JPsi events. A similar qualitative behavior is observed in photoproduced \JPsi events in $\gamma+p$ interactions by ALICE~\cite{TheALICE:2014dwa} where the ratio of non-exclusive to exclusive \JPsi events is found to be larger at high-$x$ with respect to that at low-$x$. Due to the small sample size of this analysis, the coherent \JPsi cross section is measured by summing up both configurations.

The combined acceptance ($A$) and efficiency ($\varepsilon$) correction factor for \JPsi events in the $\Xnzn$ break-up mode, $(A\,\varepsilon)^{\JPsi}$, is $5.9 \pm 0.5\syst$\%. The 8\% systematic uncertainty on the corrections are described in Section~\ref{sec:Uncert}. Two factors contribute to the $(A\,\varepsilon)^{\JPsi}$: 1) the product of acceptance multiplied by the offline reconstruction efficiency and 2) the trigger efficiency ($\varepsilon_{\text{trig}}$). The first term is measured to be $12.0 \pm 0.5\syst$\%. It is obtained from both data and MC simulations. The \STARLIGHT generator is used as an input to the full \GEANTfour~\cite{Agostinelli:2002hh} simulation of the CMS detector. This simulation is used to model the efficiency for all of the selections except the HF and the muon quality requirements. Zero bias data are used to compute the efficiency of the HF requirement, while the UPC data are used to compute the efficiency of the muon quality requirements. The offline selection discussed above is applied, but the trigger requirement is not demanded at this stage of the efficiency calculation. The UPC trigger efficiency $\varepsilon_\text{trig}$ for events passing the event selection is $49.5\pm 3.5\syst$\%. This is computed by taking the product of the efficiencies of the individual components: $\varepsilon_{\text{trig}}=\varepsilon_{\mathrm{ZDC}}\, \varepsilon_{\text{pixel-track}}\,\varepsilon_{\mathrm{BSC}}\,\varepsilon_{\text{dimuon}}$. Because these trigger components are uncorrelated to each other they are measured separately. The $\varepsilon_{\text{dimuon}}$ term is measured to be $0.71 \pm 0.02\syst$ from the analysis of the UPC data using the ``tag-and-probe" method~\cite{Chatrchyan:2012xi} in which coherent \JPsi candidates are reconstructed for a wider kinematic range than in the analysis. Two different methods to compute $\varepsilon_{\text{dimuon}}$ are studied corresponding to two different background parametrizations. Since both methods give consistent results within the statistical uncertainty, the $\varepsilon_{\text{dimuon}}$ systematic uncertainty is found to be at the 2--3\% level. The other components of the trigger efficiency do not require the reconstruction of coherent \JPsi candidates and they are measured separately using control triggers: $\varepsilon_{\mathrm{ZDC}}= 0.91\pm0.03\syst$, $\varepsilon_{\text{pixel-track}}= 0.76 \pm0.03\syst$, and $\varepsilon_{\mathrm{BSC}}$ is fully efficient. The systematic uncertainty for the acceptance and efficiency correction is discussed in the following section.

\section{\label{sec:Uncert} Systematic uncertainties and cross-checks}
The systematic uncertainties are summarized in Table~\ref{tab:sumsyst} and can be divided into three groups. The first group corresponds to the systematic uncertainty due to the signal extraction (5\%). The second group corresponds to the acceptance times efficiency correction (8\% after combining the uncertainties on the neutron detection efficiency, HF energy requirement, MC correction, ZDC trigger efficiency, and \JPsi reconstruction efficiency). The third group corresponds to the uncertainty in the luminosity determination (5\%) and in the branching ratio (0.55\%). The individual uncertainties are summarized below.

\begin{enumerate}

\item The uncertainty in the signal extraction is found to be 5\%. To estimate this uncertainty, the fitting functions used to describe the invariant mass distribution of the \JPsi and the continuum are changed to a Gaussian or Landau distribution, respectively. Also the mass region used for the signal extraction is changed to $2.4<m(\PGmp\PGmm)<8.0$\GeV. The systematic uncertainty is provided by the maximum variation of the results.

\item The uncertainty in the neutron detection efficiency is found to be 6\%. This uncertainty is mainly due to the presence of low-frequency noise in the readout and is estimated by comparing results from two different reconstruction algorithms. For each event the ZDC signal is recorded in 10 time slices of 25\unit{ns} each. The standard reconstruction method uses the difference between the signal in the main time slice and the following one. This differentiation suppresses the low-frequency noise. The alternative method estimates the noise from time slices before the main signal.

\item The uncertainty associated with the HF energy requirement is found to be 2\%. To estimate this uncertainty, the HF energy limit is decreased from 3.85 to 2.95\GeV, changing the limit from keeping 99\% of the electronic noise events to 95\%. Also, the definition of the HF energy requirement is varied by using the signal from groups of calorimeter cells known as towers, instead of the individual cells. The $\eta$ symmetry of the calorimeters is checked by defining separate limits for HF+ and HF$-$ for both individual cells and towers. The analysis is repeated for each case and the root-mean-square of the final number of signal candidates is used to estimate the systematic uncertainty associated with this requirement.

\item The uncertainty in the MC acceptance corrections is found to be 1\%. This is estimated by varying the \pt and rapidity shapes ($\pm$30\% away from the mean distribution) used to produce these corrections. As shown in Section~\ref{sec:DataAndAnalysis2}, \STARLIGHT reproduces very well the \pt shape for the various processes. The shape of the \pt distributions reflects the nuclear density distribution, which has little uncertainty.

\item The uncertainty for the ZDC component of the UPC trigger is found to be  3\%. This is estimated by using dedicated monitoring triggers.

\item The uncertainty for the \JPsi reconstruction efficiency is found to be 4\%. This is computed using the track reconstruction efficiency uncertainty that is found to be 1--2\%~\cite{CMS:2010mua}.

\item The uncertainty of the integrated luminosity determination is estimated to be 5\%, based on the analysis of data from van der Meer scans~\cite{vdmanalysis}. This uncertainty also covers the possible multiple interactions in the same bunch crossing originating from electromagnetic dissociation (EMD) processes which could affect the exclusivity requirement.

\item The uncertainty in the branching fraction for \JPsi decay into muons is 0.55\%~\cite{Agashe:2014kda}.

\item A contamination from an electromagnetic $e^{+}e^{-}$ pair could cause a possible loss of events, where one of the electrons hits the BSC scintillator and thus vetoes the event. Using a control data sample where no veto at the trigger level is applied, an upper limit on such an inefficiency is found by the ALICE collaboration to be smaller than 2\% in the coherent \JPsi analysis, at forward rapidity~\cite{Abelev:2012ba}. Since no data sample, with a comparable luminosity to the one used in this analysis, exist without a veto on the BSC, and in order to be conservative, a 2\% systematic uncertainty is assigned due to possible contamination from two-photon $e^{+}e^{-}$ background.

\end{enumerate}

 These individual systematic uncertainties are added in quadrature resulting in a total systematic uncertainty of { 11\%} for the coherent \JPsi cross section in the $\Xznz$ configuration.

As an additional cross-check of the overall analysis, the $\gamma +  \gamma$ process is studied. As discussed in Section~\ref{sec:DataAndAnalysis2}, the resulting yield of $\gamma + \gamma$ events in the $2.6<m(\PGmp\PGmm)<3.5$\GeV mass interval is $N^{\gamma + \gamma}_{\Xznz} = 75.2 \pm 12.7\stat\pm8.3\syst$, while the measured cross section is $44.2\pm1.8\stat\pm0.40\syst\unit{$\mu$b}$. This result is consistent with the QED calculation provided by the \STARLIGHT MC at the one standard deviation level. The $\gamma + \gamma \to \PGmp\PGmm$ cross section in the dimuon mass range 4 to 8\GeV (not shown) is also found to be in agreement with the \STARLIGHT prediction within one standard deviation, when considering the statistical and systematic uncertainties.

\begin{table}[htb]
\centering
\topcaption{Summary of systematic uncertainties for coherent \JPsi events in the $\Xznz$ configuration.}
\label{tab:sumsyst}
\begin{tabular}{lc}
\hline
Source & Uncertainty  \\
\hline
{(1) Signal extraction} & 5\%    \\
 (2) Neutron tagging & 6\%  \\
(3) HF energy limit & 2\%    \\
(4) MC acceptance corrections & 1\%    \\
(5) ZDC efficiency estimation & 3\%    \\
(6) Tracking reconstruction & 4\%    \\
(7) Int. luminosity determination & 5\% \\
(8) Branching fraction & 0.55\%\\
(9) Two-photon $\Pep\Pem$ background & 2\%\\
\hline
Total & {11\%}    \\
\hline
\end{tabular}
\end{table}
\section{\label{sec:Results} Results and comparison to theoretical models on photonuclear interactions}
For the $\Xznz$ break-up mode, the coherent \JPsi cross section in the dimuon decay channel is given by
\begin{equation}
\frac{\rd\sigma^{\text{coh}}_{\Xznz}}{\rd y}(\JPsi)   = \frac{N^{\text{coh}}_{\Xznz}}{\mathcal{B} (\JPsi \to \PGmp\PGmm) \, \mathcal{L}_\text{int} \, \Delta y \, (A\,\varepsilon)^{\JPsi} }
\end{equation}
where $\mathcal{B} (\JPsi \to \PGmp\PGmm)= 5.96\pm0.03\syst$\% is the branching fraction of \JPsi to dimuons~\cite{Agashe:2014kda}, $N^{\text{coh}}_{\Xznz}$ is the coherent \JPsi yield of prompt \JPsi candidates for $\pt<0.15\GeV$, $\mathcal{L}_\text{int}=159\pm8\syst\mubinv$ is an integrated luminosity, $\Delta y=1$ is the rapidity bin width, and $(A\,\varepsilon)^{\JPsi}=5.9 \pm 0.5\stat$\% is the combined acceptance times efficiency correction factor as discussed in Section~\ref{sec:DataAndAnalysis2}.
The coherent \JPsi yield of prompt \JPsi candidates is given by
\begin{equation}
N^{\text{coh}}_{\Xznz} = \frac{N_{\text{yield}}}{1+f_{{D}}}
\end{equation}
where $N_{\text{yield}}$ is the coherent \JPsi yield as extracted from the fit shown in Fig.~\ref{Figure1}, and $f_{D}$ is the fraction of \JPsi mesons coming from coherent $\Pgy\to\JPsi + \text{anything}$. As mentioned in Section~\ref{sec:DataAndAnalysis2}, $N_{yield} =  207 \pm 18\stat$ for coherent \JPsi candidates with $\pt<0.15\GeV$ in the rapidity interval $1.8<\abs{y}<2.3$. There are not enough data to perform a coherent $\Pgy$ analysis, so the feed-down correction has to rely on MC simulations. In order to calculate $f_{D}$, coherent $\Pgy$ events are simulated using \STARLIGHT, while \PYTHIA is used to simulate the $\Pgy$ decay into the \JPsi ~\cite{Abelev:2012ba, Abbas:2013oua} obtaining $f_{D}= 0.018\pm0.011\thy$. The theoretical uncertainty of 60\% in $f_{D}$ is obtained from~\cite{Abelev:2012ba, Abbas:2013oua}. The resulting coherent \JPsi yield for prompt \JPsi candidates is $N^{\text{coh}}_{\Xznz} = 203 \pm 18\stat$. Thus, the coherent \JPsi photoproduction cross section for prompt \JPsi mesons in the $\Xnzn$ break-up mode is ${\rd\sigma^{\text{coh}}_{\Xznz}}/{\rd y}(\JPsi)  = 0.36\pm 0.04\stat\pm0.04\syst\unit{mb}$.

Although the ${\rd\sigma^{\text{coh}}_{\Xznz}}/{\rd y}(\JPsi)$ measurement is interesting in its own right~\cite{Guzey:2016piu,Rebyakova:2011vf}, it is also relevant to compare our results to the theoretical predictions and recent results from the ALICE collaboration~\cite{Abelev:2012ba, Abbas:2013oua} that are available for the total coherent \JPsi cross section. As mentioned in Section~\ref{sec:DataAndAnalysis1}, one of the UPC trigger requirements is the presence of at least one forward neutron. For this reason it is not possible to scale the measured coherent \JPsi cross section in the $\Xnzn$ break-up mode to the total cross section using our own data. However, as mentioned in Section~\ref{sec:DataAndAnalysis1}, \STARLIGHT can simulate coherent vector meson photoproduction in the various break-up modes for one or both Pb nuclei. The \STARLIGHT MC generator is found to give a good description of the break-up ratios on coherent $\rho^{0}$ photoproduction measured by STAR~\cite{Abelev:2007nb} and ALICE~\cite{Adam:2015gsa}. It is also found to give a good description of the fraction of coherent \JPsi events with no neutron emitted with respect to the total number of coherent \JPsi events, measured by ALICE~\cite{Abbas:2013oua}. Moreover, \STARLIGHT gives a good description of the break-up ratios measured in this analysis. We measure the ratios of the coherent \JPsi cross section in two different break-up modes ($\XnXn$ and 1$_{n}$1$_{n}$) to that of the $\Xnzn$ mode for \JPsi events with $\pt<0.15\GeV$ and in the rapidity interval $1.8<\abs{y}<2.3$. The measured break-up ratios are $0.36\pm0.04\stat$ for $\XnXn /\Xnzn$ and $0.03\pm0.01\stat$ for 1$_{n}$1$_{n}/\Xnzn$, while the \STARLIGHT prediction is $0.37 \pm0.04\thy$ and $0.020 \pm 0.002\thy$, respectively. These ratios are also compatible with the extracted \JPsi yield for each break-up configuration, determined with the signal extraction procedure described in Section~\ref{sec:DataAndAnalysis2}. Only statistical uncertainties in the measured break-up ratios are given since these dominate over the systematic uncertainties. The feed-down correction from $\Pgy$ decays is not applied for these ratios since this contribution is expected to cancel out in the ratio. The 10\% uncertainty quoted in the \STARLIGHT prediction for the break-up mode scaling factors is based on recent results on UPC $\rho^{0}$ photoproduction from the ALICE collaboration~\cite{Adam:2015gsa}. Note that the neutron break-up theoretical description is independent of whether a \JPsi or a $\rho^{0}$ is produced~\cite{Adam:2015gsa,Baltz:2009jk}. The scaling factor between the $\Xznz$ break-up mode and the total cross section is $5.1\pm 0.5\thy$. After applying this scaling factor we obtain the total coherent \JPsi photoproduction cross section ${\rd\sigma^\text{coh}}/{\rd y}(\JPsi)  = 1.82\pm0.22\stat\pm0.20\syst\pm0.19\thy\unit{mb}$.

\begin{figure}[htb]
\centering
\includegraphics[width=\cmsFigWidth]{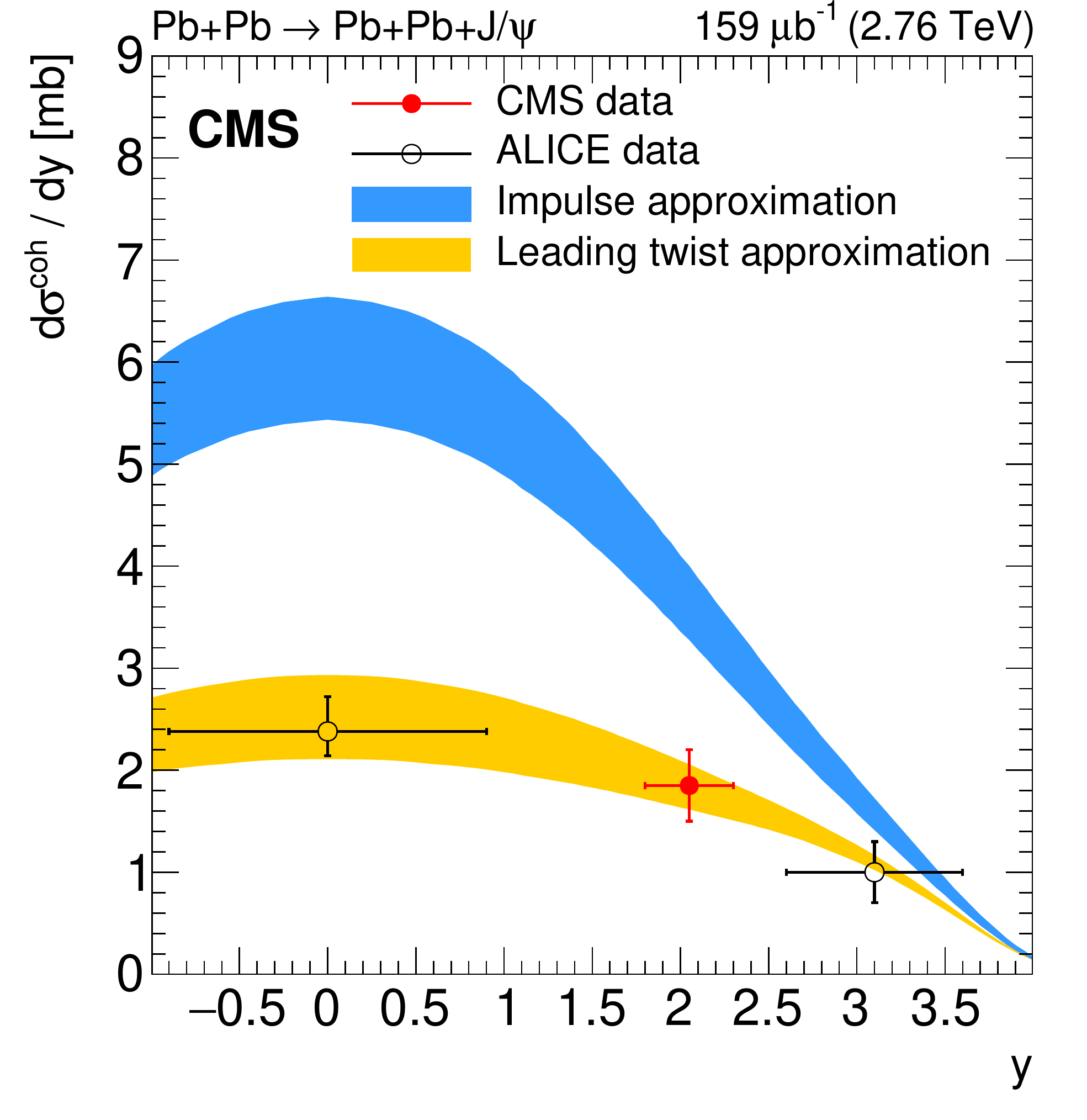}
   \caption{\label{fig5} Differential cross section versus rapidity for coherent \JPsi production in ultra-peripheral PbPb collisions at $\sqrt{s_{NN}}=2.76$\TeV, measured by ALICE~\cite{Abelev:2012ba, Abbas:2013oua} and CMS (see text for details). The vertical error bars include the statistical and systematic uncertainties added in quadrature, and the horizontal bars represent the range of the measurements in $y$. Also the impulse approximation and the leading twist approximation calculations are shown (see text for details).}
\end{figure}

In Fig.~\ref{fig5}, the coherent \JPsi photoproduction cross section is compared to recent ALICE measurements~\cite{Abelev:2012ba, Abbas:2013oua}, to calculations by~Guzey \etal~\cite{Guzey:2013jaa,Guzey:2013xba} based on the impulse approximation, and to results obtained using the leading twist approximation (see below). The data from ALICE and CMS show a steady decrease with rapidity.

The leading twist approximation prediction is obtained from Ref.~\cite{Guzey:2013jaa} and is in good agreement with the data. It is a calculation at the partonic level that uses a diffractive proton PDF as an input, following the leading twist approximation which is based on a generalization of the Gribov--Glauber nuclear shadowing approach~\cite{Gribov:1968jf}. The theoretical uncertainty band for the leading twist approximation result shown in~Fig.~\ref{fig5} is 12\% and is due to the uncertainty in the strength of the gluon recombination mechanism. This uncertainty is uncorrelated with the photon flux uncertainty. The nuclear gluon distribution uncertainty is largest at mid-rapidity where $x \sim 10^{-3}$ in the nuclear gluon distribution. At forward rapidity, integrating over all possible emitted neutron configurations, there is a two-fold ambiguity about the photon direction. In this region, the measurements are mostly sensitive to $x \sim 10^{-2}$~\cite{Abelev:2012ba}.

The data are also compared to the impulse approximation result that uses data from exclusive \JPsi photoproduction in $\gamma$ + p interactions to estimate the coherent \JPsi cross section in $\gamma+ \Pb$ collisions. The impulse approximation calculation neglects all nuclear effects such as the expected modification of the gluon density in the lead nuclei compared to that of the proton. This calculation overpredicts the CMS measurement by more than 3 standard deviations in the rapidity interval $1.8<\abs{y}<2.3$, when adding the experimental and theoretical uncertainties in quadrature.

The cross section for vector meson photoproduction in ultra-peripheral PbPb collisions is given by the sum of two cross section terms, since photons can be emitted by either of the colliding Pb nuclei. Each term is the product of three quantities: the photon flux, the integral over squared nuclear form factor $F_{A}(t)$ and the forward differential cross section $\rd\sigma/\rd t(t=0)$ of $\gamma + \Pp \to\JPsi+ \Pp$, where $t$ is the momentum transfer from the target nucleus squared. The $F_{A}(t)$ is the Fourier transform of the matter density $\rho(t)$, while the elementary cross section ${\rd\sigma}/{\rd t}$ has been measured by various collaborations~\cite{Aktas:2005xu,Chekanov:2002xi,Aaltonen:2009kg,TheALICE:2014dwa,Aaij:2014iea}, as described in Section~\ref{sec:Intro}. The impulse approximation result shown in Fig.~\ref{fig5} is performed by Guzey \etal using the methods they describe in Ref.~\cite{Guzey:2013xba} with a pQCD motivated parametrization~\cite{Guzey:2013qza} of exclusive \JPsi data in $\gamma+ \Pp$ interactions which incorporates very recent LHC results~\cite{TheALICE:2014dwa,Aaij:2014iea}. Thus, in the impulse approximation there is an experimental uncertainty associated to fitting the measured elementary cross section data to the parametrization~\cite{Guzey:2013qza} and this uncertainty is at the 4\% level for the relevant photon-proton center-of-mass energies discussed in this analysis. In addition, there are two theoretical uncertainties in the impulse approximation calculation. The first theoretical uncertainty is due to the matter density distribution and is estimated to be 5\% based on studies of several matter distribution densities~\cite{Guzey:2013xba}. The second theoretical uncertainty is due to the uncertainty in the photon flux and is estimated to be 5\%. This is dominated by the treatment of the photon flux factor for the case when the PbPb collisions take place at small impact parameters ${\sim}2 R_{A}$. These two uncertainties are correlated and so to be conservative the combined theoretical uncertainty is taken to be 10\%.

The data are also consistent with the central value of the EPS09 global fit from 2009 (not shown), which has large uncertainties~\cite{Eskola:2009uj}. Other calculations of the coherent \JPsi cross section are not considered because the theoretical uncertainties are not available.

\section{\label{sec:Conc} Summary}
The coherent \JPsi photoproduction cross section in ultra-peripheral PbPb collisions at $\sqrt{s_{NN}}$ = 2.76~TeV, in conjunction with at least one neutron on one side of the interaction point and no neutron activity on the other side, is measured to be ${\rd\sigma^{\text{coh}}_{\Xznz}}/{\rd y}(\JPsi)  = 0.36\pm0.04\stat\pm 0.04\syst\unit{mb}$ in the rapidity interval $1.8<\abs{y}<2.3$. This measurement is extrapolated to the total coherent \JPsi cross section, resulting in ${\rd\sigma^{\text{coh}}}/{\rd y}(\JPsi)  = 1.82\pm0.22\stat\pm0.20\syst\pm0.19\thy\unit{mb}$ in the measured rapidity interval. These results complement recent measurements on coherent \JPsi photoproduction in ultra-peripheral PbPb collisions at $\sqrt{s_{NN}}$ = 2.76\TeV by the ALICE collaboration. An impulse approximation model prediction is strongly disfavored, indicating that nuclear effects expected to be present at low $x$ and $Q^{2}$ values are needed to describe the data. The prediction given by the leading twist approximation, which includes nuclear gluon shadowing, is consistent with the data. In addition, we observe that, in contrast to coherent \JPsi events, the vast majority of incoherent \JPsi candidates are in the configuration when the \JPsi and the emitted neutrons are in the same rapidity hemisphere (high-$x$ component). This is qualitative agreement with the recent photoproducted \JPsi analysis off protons by the ALICE collaboration.

\begin{acknowledgments}
We congratulate our colleagues in the CERN accelerator departments for the excellent performance of the LHC and thank the technical and administrative staffs at CERN and at other CMS institutes for their contributions to the success of the CMS effort. In addition, we gratefully acknowledge the computing centers and personnel of the Worldwide LHC Computing Grid for delivering so effectively the computing infrastructure essential to our analyses. Finally, we acknowledge the enduring support for the construction and operation of the LHC and the CMS detector provided by the following funding agencies: BMWFW and FWF (Austria); FNRS and FWO (Belgium); CNPq, CAPES, FAPERJ, and FAPESP (Brazil); MES (Bulgaria); CERN; CAS, MoST, and NSFC (China); COLCIENCIAS (Colombia); MSES and CSF (Croatia); RPF (Cyprus); MoER, ERC IUT and ERDF (Estonia); Academy of Finland, MEC, and HIP (Finland); CEA and CNRS/IN2P3 (France); BMBF, DFG, and HGF (Germany); GSRT (Greece); OTKA and NIH (Hungary); DAE and DST (India); IPM (Iran); SFI (Ireland); INFN (Italy); MSIP and NRF (Republic of Korea); LAS (Lithuania); MOE and UM (Malaysia); BUAP, CINVESTAV, CONACYT, LNS, SEP, and UASLP-FAI (Mexico); MBIE (New Zealand); PAEC (Pakistan); MSHE and NSC (Poland); FCT (Portugal); JINR (Dubna); MON, RosAtom, RAS and RFBR (Russia); MESTD (Serbia); SEIDI and CPAN (Spain); Swiss Funding Agencies (Switzerland); MST (Taipei); ThEPCenter, IPST, STAR and NSTDA (Thailand); TUBITAK and TAEK (Turkey); NASU and SFFR (Ukraine); STFC (United Kingdom); DOE and NSF (USA).

Individuals have received support from the Marie-Curie program and the European Research Council and EPLANET (European Union); the Leventis Foundation; the A. P. Sloan Foundation; the Alexander von Humboldt Foundation; the Belgian Federal Science Policy Office; the Fonds pour la Formation \`a la Recherche dans l'Industrie et dans l'Agriculture (FRIA-Belgium); the Agentschap voor Innovatie door Wetenschap en Technologie (IWT-Belgium); the Ministry of Education, Youth and Sports (MEYS) of the Czech Republic; the Council of Science and Industrial Research, India; the HOMING PLUS program of the Foundation for Polish Science, cofinanced from European Union, Regional Development Fund; the Mobility Plus program of the Ministry of Science and Higher Education (Poland); the OPUS program of the National Science Center (Poland); the Thalis and Aristeia programs cofinanced by EU-ESF and the Greek NSRF; the National Priorities Research Program by Qatar National Research Fund; the Programa Clar\'in-COFUND del Principado de Asturias; the Rachadapisek Sompot Fund for Postdoctoral Fellowship, Chulalongkorn University (Thailand); the Chulalongkorn Academic into Its 2nd Century Project Advancement Project (Thailand); and the Welch Foundation, contract C-1845.
\end{acknowledgments}
\bibliography{auto_generated}

\cleardoublepage \appendix\section{The CMS Collaboration \label{app:collab}}\begin{sloppypar}\hyphenpenalty=5000\widowpenalty=500\clubpenalty=5000\textbf{Yerevan Physics Institute,  Yerevan,  Armenia}\\*[0pt]
V.~Khachatryan, A.M.~Sirunyan, A.~Tumasyan
\vskip\cmsinstskip
\textbf{Institut f\"{u}r Hochenergiephysik der OeAW,  Wien,  Austria}\\*[0pt]
W.~Adam, E.~Asilar, T.~Bergauer, J.~Brandstetter, E.~Brondolin, M.~Dragicevic, J.~Er\"{o}, M.~Flechl, M.~Friedl, R.~Fr\"{u}hwirth\cmsAuthorMark{1}, V.M.~Ghete, C.~Hartl, N.~H\"{o}rmann, J.~Hrubec, M.~Jeitler\cmsAuthorMark{1}, A.~K\"{o}nig, M.~Krammer\cmsAuthorMark{1}, I.~Kr\"{a}tschmer, D.~Liko, T.~Matsushita, I.~Mikulec, D.~Rabady, N.~Rad, B.~Rahbaran, H.~Rohringer, J.~Schieck\cmsAuthorMark{1}, R.~Sch\"{o}fbeck, J.~Strauss, W.~Treberer-Treberspurg, W.~Waltenberger, C.-E.~Wulz\cmsAuthorMark{1}
% ...
\vskip\cmsinstskip
% ...
1:~~Also at Vienna University of Technology, Vienna, Austria\\
% ...

\end{sloppypar}
\end{document}